\documentclass[12pt,preprint]{aastex}
\begin{document}

\title{ UV Continuum, Physical Conditions and Filling Factor
in Active Galactic Nuclei}

\author{ Lucimara P. Martins\altaffilmark{1}}
\affil{Space Telescope Science Institute, 3700 San Martin Drive
Baltimore, MD 21218}
\email{martins@stsci.edu}

\author{Sueli M. Viegas \& Ruth Gruenwald}
\affil{Instituto de Astronomia, Geof\' \i sica e
Ci\^encias Atmosf\'ericas, S\~ao Paulo, Brazil}

\altaffiltext{1}{Instituto de Astronomia, Geof\' \i sica e
Ci\^encias Atmosf\'ericas, S\~ao Paulo, Brazil}

\begin{abstract}
The narrow line region of active galaxies is formed by gas clouds
surrounded by a diluted gas.
Standard one-dimensional photoionization models are usually used to model this region
in order to reproduce the observed emission lines. Since the
narrow line region is not homogeneous,
 two major types of models are used: (a) those assuming a
 homogeneous gas distribution and a  filling factor less than unity to mimic the
presence of the emitting clouds; (b) those based on
a composition of single-cloud models
combined in order to obtain the observed spectra. The first method is largely used
but may induce to misleading conclusions as shown in this paper.
The second one is more appropriate,
but requires a large number of observed lines in order to limit the number of
single models used. After discussing the case of an extragalactic HII region,
for which the ionizing radiation spectrum is better known,
we show that 1-D  models for the narrow line region
 with a filling factor less than unit
do not properly mimic the clumpiness, but just simulates an overall
lower density. Multi-cloud models lead to more reliable results.
Both models are tested in this paper, using the emission-line
spectra of two well-known Seyfert galaxies, NGC 4151 and NGC 1068.
It is shown that ionizing radiation spectra with a blue bump cannot
be excluded by multi-cloud models, although excluded by Alexander et al. (1999, 2000)
using homogeneous models with a filling factor less than unity.

\end{abstract}


\section{Introduction}

The intrinsic spectral energy distribution (SED) of active galactic nuclei
(AGN)
extends from radio up to $\gamma$ rays with observations available
in a wide wavelength range. However, due to Galactic and intrinsic absorption, the
SED above the Lyman limit cannot be observed. Since the main mechanism powering the
emission-line spectrum of AGN is photoionization, it is important to
know the shape of the UV radiation. As suggested by the
observed continuum, it is usually assumed that the SED from ultraviolet to
soft X-rays is basically  a  power law with at least two other
components corresponding to the so-called ``big blue bump'',
usually represented by a blackbody radiation with temperature between 1.5
to 3.0 $\times$ 10$^5$ K (see, for instance Oliva et al. 1994, Prieto \& Viegas 2000),
and to the soft X-ray excess, frequently described as a thermal component
 (see, for instance, Terashima et al. 2002).
 In addition, the  SED is also important to test the emission
by  supermassive black holes and to discriminate among
 different models (Narayan, Kato \& Homna
 1997), as well as to help finding the signature of shocks in the
narrow emission-line
region (Contini, Viegas e Prieto 2002, and references therein).

Many observations of Seyfert galaxies point to the existence of three
kinematically distinct emission regions:
(a) an unresolved broad line region (BLR), with typical line widths of
$\ge$ 5000 km s$^{-1}$, densities above
10$^8$ cm$^{-3}$ and ionization parameter U $\approx$ 0.03-1; (b) a narrow
line region (NLR), with lower densities ($\leq$ 10$^7$ cm$^{-3}$)
and lower velocities ($\leq$ 500 km s$^{-1}$)
gas clouds with U in the range 0.03 to 0.1 embedded in a more diluted gas;
(c) an extended emission-line region
(EELR) with typical line widths of less than 50 km s$^{-1}$, densities
below 10$^3$ cm$^{-3}$, and U smaller than 0.005
(Schulz and Komossa 1993).

The emission-line spectra of AGN show a rich variety of
optical and UV lines,
including those from highly ionized species (coronal lines).
The observed line ratios point  to photoionization  as the
dominant mechanism powering  these objects.
Recently, however, near infrared observations of narrow-line Sy 1 galaxies
tend to favor shocks as the mechanism producing coronal lines (Rodriguez-Ardila
et al. 2002).
Nevertheless, photoionization being the main mechanism contributing to the
emission-line spectra of AGN,
photoionization models can be used to constrain the SED by fitting the
observed lines.

Photoionization codes presently available were developed in the 70's.
They are one-dimensional (1-D) codes that
 usually assume a homogeneous gas distribution with
a well-defined symmetry (plan-parallel or spherically symmetric).
The standard input parameters are the shape and intensity
of the ionizing radiation spectrum, the density of the gas,
and the elemental abundances.
Therefore,  1-D models are extremely limited
to simulate clumpy regions like HII regions or the
narrow line region of AGN.

In order to simulate an inhomogeneous region, two types of models
are usually proposed: (a) those assuming a
homogeneous distribution of gas and  a filling factor
less than unity (see, for instance Osterbrock 1989);
(b) a multi-cloud model  generated by a weighted average
emission-line spectrum from  single-cloud simulations (with $\epsilon$ = 1),
as firstly proposed by Viegas-Aldrovandi \& Gruenwald (1988),
and more recently used by other authors
(Baldwin et al. 1995, Binette, Wilson \& Storchi-Bergmann 1996,
Fergunson et al. 1997, Komossa \& Shulz 1997,
Kraemer \& Crenshaw 2000). Multi-cloud models of the NLR, based
on sigle-cloud models which take into account the coupled
effect of photoionization and shocks, have also been used to
analyze the NLR of AGN (Contini,
Prieto  \& Viegas 1998a, 1998b, Contini \& Viegas 2001,
Contini et al. 2002).

The idea of a filling factor  was first introduced by Osterbrock \&
Flatter (1959) in order to explain the radio brightness temperature of
HII regions lower than expected, indicating that the gas has a clumpy
distribution. The filling factor $\epsilon$ is the fraction of the total
volume occupied by the clumps or condensations. 
Since 1-D photoionization codes are unable to
properly deal with an inhomogeneous gas distribution, the optical
depth of each shell across the nebula is multiplied by the
chosen filling factor (see, for instance, Osterbrock 1989).
 The net effect corresponds to a decrease of
the average density, increasing the size of the ionized region,
particularly the low ionization one, as it will be discussed below.

In the following, the expression {\it `model with a filling factor'}
refers to a model with a filling factor less than unity. Because
of the increase of the low ionization zone relative to the
high ionization zone, this kind of model tends to overestimate
the low-ionization lines. Therefore, when simulating the NLR
we must keep in mind this systematic effect and how it
can affect the conclusions reached

A multi-cloud model offers a more realistic view of the NLR, as long as
a proper choice of single-clouds (densities and ionization parameters)
is achieved. Assuming that the chemical composition
is the same, each group of single-clouds contributing to the
emission-line spectrum is characterized by
the density and ionization parameter (ratio of ionizing photons to gas
density). The simulated emission-line spectrum corresponds to the weighted
average of the line intensities coming from each group.
The quality of the fit is related to the number of observed emission-lines.
A better fit to the observed spectrum is obtained when a larger number
of observed lines is available, allowing a fine tuning of the physical
conditions producing the lines, leading to a better choice of the groups
of single clouds.

Recently, NLR photoionization models have been used to constrain the
shape of the spectral energy distribution of active nuclei
(Oliva et al. 1994, Alexander et al. 1999 and 2000,
Prieto \& Viegas  2000).
Here, our goal is to discuss the dependence of the SED on the type of model
used to describe the NLR. Before analyzing the emission-line spectrum of two Seyfert
galaxies (NGC 4151 and NGC 1068),  we discuss the
filling factor problem related to photoionization models for extragalactic
HII regions. For these objects the ionizing radiation spectrum is better
known and the chemical composition obtained by empirical methods is reliable.
Photoionization simulations of the NW knot of I Zw 18 provide
a good example of the effect of a filling factor on the gas temperature
and  how models with a filling factor may induce to misleading results.

In \S 2, models for an extragalactic
HII region are used to illustrate the effect of the
filling factor, and a model for the NW knot of I Zw 18
is discussed. In \S 3, the observational data for NGC 1068 and NGC 4151
are presented, whereas photoionization models appear in \S 4.
 The concluding remarks are presented in \S 5.

\section{Filling Factor Models}

The effect of $\epsilon$  $<$ 1 in 1-D models is to
increase the size of the low-ionization zone,
consequently increasing the ionized region and
reducing the average temperature of the ionized gas.
Models  with $\epsilon$ = 1 should provide a higher average
gas temperature, increasing the calculated [O III] line ratio.
In order to illustrate the effect of a filling factor  less
than unit, the results of HII region models
with $\epsilon$  $\leq$ 1, obtained by the 1-D code
Aangaba (Gruenwald \& Viegas 1992), are discussed.

We adopt the
ionizing radiation of a zero age stellar cluster from Cid-Fernandes
et al. (1992) shown in Figure 1 (top panel),
 with a number of ionizing photons Q$_H$ = 10$^{51}$
s$^{-1}$, a uniform distribution of the gas with hydrogen density
equal to 100  cm$^{-3}$, and chemical abundances of I Zw 18
(Izotov \& Thuan 1998).
The results obtained for the outer radius of the
HII region, as well as  the average
temperature T$_{OIII}$, as a function of $\epsilon$, are
shown in Figure 1 (middle panel), while  the ionic fractions of
O$^0$, O$^+$ and O$^{++}$ appear in Figure 1 (bottom panel).
As expected, the outer radius, O$^0$ and  O$^+$  increase with decreasing
 $\epsilon$, while  T$_{OIII}$ and O$^{++}$ decrease,
 corresponding to the expansion of the low-ionization zone
for smaller values of $\epsilon$.

The shape of the ionizing radiation is another factor
defining the relative size of the high-ionization
to the low-ionization zone. To exemplify this point
the zero age stellar cluster continuum spectrum was
modified between 1 Ryd and 4 Ryd (Fig. 1, top panel)
and the results are also shown in Figure 1 (middle and
bottom panels).
 The filling factor effect is similar for
 the two types of ionizing spectra. However,
the ionic fractions of O$^+$ and O$^{++}$ are larger
while the O$^0$ fraction is lower
when the spectrum has less photons near the Lyman limit.
Notice  that, in this case,  O$^{++}$/O increases by a larger
amount. Ions of other elements follow the same behavior.
Therefore, the increase of the low-ionization zone due
to a filling factor less than unity, leading
to stronger low-ionization emission-lines, can be
compensated by adopting an ionizing radiation with
less photons near the Lyman limit.

\subsection{The Star Forming Galaxy I Zw 18}

Recently, 1-D photoionization models for the NW knot of
I Zw 18, a metal poor extragalactic HII region,
have been used by Stasinska \& Schaerer (1999, hereafter SS99)
to analyze its emission-line spectrum.
They concluded that in addition to photoionization,
another heating mechanism is necessary in order to explain all
observed features, i. e., the size of the ionized region, the
H$\beta$ luminosity and the emission-line ratios of the more
significant lines.

In order to model the NW knot, SS99 adopted a homogeneous gas
distribution with density equal to 100 cm$^{-3}$, derived from
the observed [S II] line ratio, as well as the chemical
abundances obtained  by Izotov \& Thuan (1998).The best fit
model should reproduce the significant emission-line ratios,
the H$\beta$ luminosity, as well as the size of the ionized region.
Adjusting the total initial mass of the stellar
cluster in order to reproduce the luminosity,  the authors
show that a homogeneous model may explain some of the line ratios, but
the calculated size of the ionized region is too small unless
a filling factor of the order of 0.01 is adopted. Their best fit model
corresponds to an homogeneous model with this filling factor, combined
with condensations simulated by higher density models.
 All the observed features are explained except for the observed
[O III] $\lambda$4363/5007 line ratio, the calculated ratio being too low.
Because this ratio is a temperature indicator,
the authors concluded that an additional heating source must
be present in the NW knot and suggested the possible presence of shocks.

Following Viegas, Contini \& Contini (1999), shocks must
be contributing to the observed emission-lines of starburst
galaxies depending on the evolutionary phase of the stellar
cluster and could be identified as an additional heating mechanism.
However, I Zw 18 is  a metal poor extragalactic
HII galaxy where the shock contribution to the physical conditions
is negligible.  Such effects could only be significant in
evolved star forming regions. In order to analyze the
apparent discrepancy between the
need for an additional heating source and the characteristics  of
I Zw 18, the SS99 model assumptions have been reviewed
(see Viegas 2002 for more details), using the code
Aangaba.

Following SS99, $\epsilon$ $<$ 1
is necessary in order to match the size of the ionized region.
However, a model with $\epsilon$ = 1  and a lower density could
reproduce it. Notice that a density of about 30 cm$^{-3}$ is
also in agreement with the observed [S II] ratio. In fact, Viegas (2002)
showed that a uniform density model for a spherical region
 with n = 30 cm$^{-3}$ and filling factor
equal to unit, photoionized by a 5.4 Myr stellar cluster,
reproduces the size of the NW knot and the H$\beta$ luminosity,
although some of the observed line ratios are not explained.
The same kind of discrepancy between observational and theoretical
line ratios was reported by SS99 concerning the homogenous model.
These authors suggested the presence of condensations to improve
the fitting. In fact, an inhomogeneous
gas distribution is suggested by the HST image of I Zw 18.
Two other models with a higher density (10$^ 4$ cm$^{-3}$),
mimicking possible condensations, are also presented by
Viegas (2002), showing that the emission-line intensities, the
H$\beta$ luminosity, as well as the size of the emission region,
are well reproduced with no need of an additional heating mechanism.

No specific combination  of dense and dilute gas is provided by SS99,
but they show that the presence of condensations improves the results for
the [O I] emission line, which was too low in the homogeneous model.
The results obtained by Viegas (2002) with dense gas
(10$^4$ cm$^{-3}$) at different distances from the stellar
cluster indicated that the best choice is a set of condensations
located in the outer zone of the ionized region. Notice that a
large fraction of H$\beta$ is emitted in the inner zone.
Taking these facts into account,
 an example of a composite model for
 the NW knot of I Zw 18 is presented in Figure 2, where
 the results for a dilute uniform gas
 (30 cm$^{-3}$, filled circle)  and a denser gas
 (10$^4$ cm$^{-3}$, filled square) are combined (circle)
 assuming that  60\% of  H$\beta$  comes from the low density
  gas and the remaining 40\% from condensations. This model
 provides a very good fit of  the line ratios, as well as of the
 size of the ionized region and of the average T$_{OIII}$.
The results presented in Fig. 2 make clear the potential of
multi-cloud  models to reproduce inhomogeneous regions without the problems
associated with models with a filling factor less than unity.
It is not our intention to claim that this is the best model
 for the NW knot of I Zw 18, but just to exemplify how an
 artificial problem created by models with filling factor can be
 solved by composite models. Notice, however, that a reliable composite
 model requires the fitting of  large number of observed lines,
 not only the most significant ones.

In brief, the adoption of a filling factor led to a ``heating problem''
and, consequently, to the suggestion of the presence of shocks, whereas
 a composite model, built from single-cloud models with different densities
and $\epsilon$ = 1, explains all the observable features without
requiring an additional heating mechanism, in agreement with the
conclusions from Viegas et al. (1999). In this case, the misleading effect of
a filling factor less than unity -- {\it artificially increasing
the low-ionization zone of the cloud and decreasing the average temperature
of the gas} -- could be easily identified because the
ionizing radiation spectrum is well known. However, for the NLR of AGN
the ``filling factor effect'' could influence the choice of
the shape of the unknown UV ionizing radiation spectrum
inferred from the models, as discussed below for two
Seyfert galaxies, NGC 1068 and NGC 4151.

\section{NGC 4151 and NGC 1068 Observational Data}

NGC 4151 is a spiral galaxy, almost face-on, with z = 0.0033 and
magnitude m$_V$ = 11.5.
Its continuum spectrum is dominated by a non-stellar
component (Kaspi et al. 1996)
which is extremely variable, as are the broad emission lines.
This galaxy is usually given as an example of  a classic Seyfert galaxy,
 but it shows a variety of physical properties of different AGN classes.
It could then be more complex than the average Seyfert
galaxies. Originally classified
as an intermediate Sy 1.5 galaxy (Osterbrock \& Koski 1976), it
presented the characteristics of a Sy 2 during its extremely
low state in 1984  (Penson \& Perez 1984). Currently NGC 4151 is
classified as a Sy 1 galaxy.

NGC 1068 is one of the brightest and closest known Sy 2 galaxies.
It is also the archetypal ``hidden Seyfert 1''  galaxy, following the
unified model for active galactic nuclei (Antonucci \& Miller 1985,
Antonucci 1993). The narrow-line region is approximately
co-spatial with a linear radio source with two
lobes (Wilson \& Ulvestad 1983).
Most of the NLR and extended emission line
region are likely powered by photoionization by a central source
(Marconi et al. 1996). Moreover, high-resolution
observations suggest kinematic
disturbance and possibly shock excitation of regions close to the
radio outflow (e.g. Axon et al. 1998).

Both objects have already been
 modeled by Alexander et al. (1999) and  Alexander et al. (2000)
(hereafter A99 and A00, respectively),
using photoionization models with a filling factor.
These authors used the observed emission-lines to
constrain the ionizing continuum spectral shape,
in particular to discuss the presence
of a blue bump. On the other hand, two-component photoionization models,
with a broken power law,  have been used  by
Kraemer \& Crenshaw (2000) to reproduce the emission-line spectra of NGC 1068.
Multi-cloud models, which take into account the
coupled effect of photoionization
by a power-law ionizing radiation and shocks,
are also discussed by Contini et al. (2002)
for NGC 4151.

In order to discuss the filling factor effect on the
inferred SED of the ionizing radiation,
NLR of these two AGN are modeled in this work. 
We will adopt the list of observed emission-lines  of NGC 4151 and NGC 1068
from A99 and A00, respectively. They provide a list of  emission-lines
 carefully selected  according to the following  criteria:
(a) only lines that are primarily formed by photoionization due to an
UV -- soft X-ray continuum from a central source are included; (b)
 all lines emitted from ions with ionization potential less than 13.6 eV
are excluded, since  these ions can be
easily ionized by other processes; (c)  only lines with reliable
 flux measurements are included, requiring that the line measurements are
not too inconsistent with the large SWS and
LWS (ISO Short- and Long-Wavelength
Spectrometer, respectively) apertures;  and finally,
(d) they excluded the [FeX] 6734 $\AA$
and [FeXI] 7892 $\AA$  coronal lines, because the collision strength
values are highly uncertain. Whenever more than one measurement for these
lines exists, they quote the average flux and use rms scatter as an
error estimate.

Notice that A99 compare the emission-line absolute fluxes
of NGC 4151 to the model results, instead of the
line intensity relative to a Balmer line.
Their argument is that the narrow  H$\beta$ flux
measurement is very uncertain,
due to the difficulty of decomposing the
 narrow and broad components. This is true indeed. Notice however that some
 of the references  regarding the observational data
present the emission lines relative to H$\beta$, without providing
 H$\beta$ absolute value (for example, Osterbrock \& Koski 1976).
 In this case, A99 uses the absolute value
from another data set. Furthermore,
some of the references are extremely old, and clearly show serious
problems with the measurements, mainly with the decomposition of
the narrow and broad line components (for example,
Oke \& Sargent 1968 where H$\alpha$/H$\beta$ = 1.61).
Because of these problems, we decided to exclude data coming from
 references where  H$\alpha$/H$\beta$ is less than 2.8,
since this clearly indicates a problem with the decomposition.
In addition, as pointed out by A99,  the narrow component of
the UV lines (Kriss et al. 1992) show  a
FWHM which is 1.5 to 3 times larger
than that of the NLR forbidden lines, which could also indicate 
a problem with the decomposition.

For NGC 1068, the absolute fluxes of the lines are provided by
the authors in all references listed by A00. However, some
 measurements seemed to be problematic as well.
Most of the observations were
made in the 70's, with photographic plates. Some of the measurements
are very discordant, which can be seen by the rms scatter
in the A00 list. Taking this into account, we decided to use only the very
strong optical
and  infrared lines, excluding  problematic
data from the average.

Therefore,  averaged line intensities in our sample
 are not exactly the same as in A99 or A00. We choose to compare
the observed and calculated line ratios, and not the absolute fluxes,
adopting  the line intensities relative to H$\beta$.
The line ratios corrected for reddening are presented
in Table 1a and 1b, respectively, assuming E(B-V) = 0.03 for NGC 4151
 and E(B-V) = 0.30  for NGC 1068, in agreement with
 those adopted by A99 and A00.
In  Tables 1a and 1b, the doublets are identified by the wavelength of
one of the lines followed by a plus sign.

\section{The Models}

The photoionization code Aangaba (Gruenwald \& Viegas 1992) is used
to model the NLR of NGC 4151 and NGC 1068,
 assuming a plan-parallel symmetry.
For each AGN, we first reproduce the results of the
best fit given by A99 and A00, using the same input parameters,
as listed in Table 2,
where U is the ionization parameter, n  the gas density,
$\epsilon$ the filling factor, and R$_i$ the distance of the gas to
the central source. Solar chemical abundances are adopted for
modeling NGC 4151. On the other hand, for NGC 1068 a low oxygen
abundance is assumed, as suggested by Netzer (1997) and adopted by
A00 for their best fit model.
From their models with a filling factor, A99 and A00
proposed that the continuum
of these objects should be represented by a
composition of power laws, as shown in
Figures 3a and 3b, for NGC 4151 and NGC 1068,
respectively. The continuum spectral shapes
above 10$^3$ eV (X-rays) and below 10 eV (optical-UV)
are obtained from observations,
while between 10 and 10$^4$ eV several different shapes have been tested.

After reproducing the results of the  best fit models
proposed by  A99 and A00 for NGC 4151 and NGC 1068, our
next step is to explain the emission-line
spectra using a multi-cloud model (based on single-cloud models with
$\epsilon$ = 1), in order to test the spectral shapes
suggested by A99 and
A00, who excluded the presence of a blue bump.

\subsection {NGC 4151 models}

As claimed by A99, the spectrum showed by a solid line in Figure 3a (spectrum A)
provides the best fit model for the observed
narrow emission-lines of NGC 4151.
Those authors did not find indications
of a big blue bump in the ionizing continuum,
 which should appear beyond the Lyman limit. On the
contrary, their best fit model suggests a trough in a power law.

As pointed out in \S 2, the trough in the ionizing spectrum
could be an artifact which  compensates for the systematic effect
 of  a filling factor less than unity in the 1-D models.  Thus,
we look for  multi-cloud models with
 another ionizing spectrum shape in order to verify if
 they are also able to reproduce the observed emission-lines.

For sake of comparison to a multi-cloud
model, we reproduce the A99 best fit model using the code Aangaba
with the input parameters listed in Table 2. Notice that our
data set is slightly different from that used by A99, as explained
above (\S 3), in particular the H$\alpha$/H$\beta$ ratio, which is
equal to 2 in their data set. In spite of the differences, the fitting
to the data is as good as that presented by A99, as
seen in Figure 4 where observed and calculated
line ratios are confronted (to be compared with Fig. 8 of A99).

Models using other spectral shapes were also obtained. Among them,
the model with spectrum B (Fig. 3a), showing a small blue bump,
also provides a good fit to the observed lines,
except for the high ionization lines,
as expected from models with such type of continuum and assuming a
filling factor less than unit. However, as in the case of the
extragalactic HII region I Zw 18, we cannot exclude this kind of spectral
shape, since the underestimation of the high ionization lines can also
be due to $\epsilon$ less than unit. In order to test this issue,
multi-cloud models are also obtained. For this
a grid of single-cloud models is built
varying the density and the ionization parameter U.

The simplest composite model includes a low- and a high-density cloud,
as adopted by Kraemer \& Crenshaw (2000)  in order to reproduce the
the optical-UV emission-line spectrum of NGC 1068.

A grid of models for  single-clouds
characterized by the density (n) and the ionization parameter (U)
was obtained.
The models were generated with these parameters in the range
10$^2$cm$^{-3}$ $\leq$ n $\leq$ 10$^5$cm$^{-3}$ and
0.005 $\leq$ U $\leq$ 4.0. To restrict the physical characteristics of the
models we used the strong lines [OIII]$\lambda$5007+ and
[MgVIII]$\lambda$ 3.03 $\micron$.
 For a given value of U and varying the density,
 we look for the best two-density models
 which can explain these emission-lines.
 Once the densities are derived, we used all
 the observed emission-line intensities (Table 1a) to
 obtain the values of U for the low and high density clouds.
 The emission-line spectrum is reproduced by a four-cloud model.
Notice that line ratios are not method dependent, i.e.,
 the same cloud composition is found if we first define
 the values of U and then the cloud densities.
The line ratios are obtained by a weighted
average from the contributions from each of the four types of cloud.
The ionization parameter,
the gas density, and the H$\beta$ flux,  as well as the
fraction of H$\beta$ (g$_i$) coming from  each type of cloud,
are listed in Table 3.
This fraction is related to the  area of the corresponding
group of clouds photoionized by the central source.
The largest contribution to the emission-line
spectrum  comes from  the low-ionization, high density
clouds.

A comparison between the observed  and calculated
emission-line ratios appears in Figure 5,
 where the results of A99 best fit
are also plotted. It is easily seen that both the A99 model,
using an ionizing SED with a trough
and $\epsilon$  $<$ 1, and our four-cloud model, using
 an ionizing radiation SED with a bump (spectrum B)
and  $\epsilon$ = 1, provide a similar fit to the observations.
This can be verified by the $\chi^2$ value, which are
2.27 and 2.28 respectively for
the A99 best fit model and our multi-cloud model.

The multi-cloud model presented above is an example of the
method used to fit the emission-lines and shows that the
ionizing spectrum may show a bump instead of a
trough as claimed by A99. In order to have a better determination
of the ionizing continuum, additional constraints,  coming from
the observed continuum spectrum from radio to X-rays, could be added
to the model, as well as the contribution of an additional
energy source, as discussed by Contini et al. (2002).

\subsection{NGC 1068 models}

In the case of NGC 1068, A00 do not find a single-component model
that satisfactorily reproduces the observed emission lines.
For this object, their best fit is a two-component model; one of the
components is calculated
with a radially varying filling factor. It is out of the scope of
this paper to reproduce such a complex model for this object,
since our goal is just to test the two approaches used
to reproduce the emission-line spectrum of an inhomogeneous region
with 1-D photoionization simulations.
  Therefore, we used input parameters of the best single-component
model of A00 to discuss these two approaches, using the two
shapes of the ionizing continuum plotted in Fig. 3b.

Spectrum A (Fig. 3b), with a trough,  was assumed in the A00 single-cloud model
which gives the best fit to the observations with
the input parameters listed in Table 2. Spectrum B is an example of
a bumpy continuum.

Following the procedure used for NGC 4151, the
emission-line spectrum of NGC 1068 is first modeled
using an uniform density distribution and a
filling factor $\epsilon$ $<$ 1. The input parameters are listed
in Table 2. The general conclusion is similar
to the one obtained for NGC 4151, i.e., regardless
of the spectral shape of the
ionizing radiation, single-cloud models with  $\epsilon$ $<$ 1
overestimate the low ionization lines with respect to the
high ionization lines, as already pointed out by A00.
The comparison between observed and calculated emission-line
intensities are shown in Figure 6 for both
types of ionizing spectra.

We now look for the best fit multi-cloud model
using the bumpy spectrum (spectrum B, Fig.3 )
in order to compare the results to those of
A00 best fit model with ionizing spectrum A and $\epsilon$ $<$ 1.
Following  the same procedure adopted for NGC 4151 multi-cloud model,
four types of clouds are defined.
The characteristics of the single clouds, as well as the
fraction of H$\beta$ coming from each type of cloud, are
presented in Table 4. The larger contribution
to the total H$\beta$ comes from the low-ionization
low-density cloud.

The comparison between observed and calculated line ratios
(Figure 7) shows that, except for  the infrared lines [NeIII] 15.56 and
[NeIII] 36.01,  our four-cloud fit, photoionized
by a bumpy spectrum,  is better than that
provided by the best A00 single-cloud model.
The corresponding $\chi^2$ value for these models are
 2.88 and 7.89, respectively.

Notice that A00 find that this simple filling factor model
was not good enough to reproduce the observed spectrum. They assumed a
more complex two-component model, obtained from
models with a filling factor. The low-density component has
a constant filling factor, while the denser one a filling
factor dependent on the distance. It is
clear that our simpler multi-cloud model provides a very good fit 
to the observed line spectrum without the problems associated to and
complexity
required by their models with a filling factor.

\section{Concluding Remarks}

The emission-line spectra of the two well-known Seyfert galaxies
NGC 4151 and NGC 1068 are used to test 1-D photoionization
 models for the NLR and to illustrate the misunderstanding
introduced by the use of models with  a filling factor less than unity to
simulate inhomogeneous regions.

The ionizing radiation of AGN is not well known,
but its shape, combined with
the filling factor as it is introduced in 1- D photoionization codes,
has a direct effect on the calculated emission lines.
In the case of the extragalactic HII regions,
 for which the ionizing continuum is better known,
 we show that a multi-cloud model can  explain
 all the observed features (size of the emitting region, H$\beta$
 luminosity and the significant emission-line ratios) without
 requiring an additional heating mechanism as inferred by SS99
 who used models with $\epsilon$ $<$ 1.

Regarding the two Seyfert galaxies
analyzed in this paper, we show that multi-cloud models
are more appropriate to simulate the NLR. They
provide a good fit to the observed emission-line spectra and
do not exclude a bump in  the SED of the ionizing radiation,
as suggested by A99 and A00.

Two-component models have been used by  Kraemer \& Crenshaw (2000) to
reproduce the optical-UV emission-line spectrum
from several resolved zones of the NLR of NGC 1068. The hydrogen
density of the dense components are in the range 2.0 $\times$ 10$^4$
to 2.5  $\times$ 10$^5$ cm$^{-3}$ and the ionization parameter
10$^{-3.8}$ $\leq$ U $\leq$ 10$^{-2.5}$ , while  for the tenuous component
the density varies from 8.0 $\times$ 10$^2$ to
2.3 $\times$ 10$^4$ cm$^{-3}$
and 10$^{-2}$ $\leq$ U $\leq$ 10$^{-1.6}$. The values of the density
and ionization parameters used in our multi-cloud model are in the range of 
those obtained for the two-cloud models proposed by Kraemer \& Crenshaw for 
several resolved zones of the NLR of NGC 1068. Notice that our model simulates
 the whole NLR and also explains the infrared lines not included in Kraemer \&
 Crhenshaw discussion.

A more comprehensive multi-cloud model of the  emission line spectrum
of NGC 4151 is presented by Contini et al. (2002).
The code used in the simulations
includes the effect of shocks in the physical conditions
of the emitting gas. The observed continuum data from radio to
X-rays are used  to constrain the model and
to find the contribution from the
different clouds to the  observed continuum and emission-line
spectra. Noticed that  the
presence of shocks is also suggested by Kraemer \& Crenshaw (2000).

In brief, in the last years  the improvement of
observational techniques,
allowing  observations with high spatial and spectral resolution,
and providing  a large amount of data in a wide range of wavelengths,
has not been followed by a similar effort to improve the interpretation
of the emitting regions with more realistic models. The
standard 1-D photoionization models are becoming inappropriate
to simulate emission-line regions like HII regions, planetary nebulae
and active galactic nuclei. The new observations
require more complex models regarding both additional  mechanisms
powering the emission (as shocks) and a better
simulation of the  geometry of the region
with a more realistic gas distribution.
It is clear that 3-D photoionization simulations would
be more appropriate to analyze the NLR of AGN, as shown in the case
of planetary nebulae (Gruenwald, Viegas \& Brogui\`ere 1997, Monteiro et al. 2000).
Such 3-D models will certainly contribute to a better discrimination
of the possible nature of the  ionizing radiation.

\begin{acknowledgements}

We are thankful to T. Alexander and H. Netzer for clarifying some
of their results and data. We thank also Claus Leitherer for the careful
reading of this paper. This paper is partially supported by
FAPESP (00/06695-0, 00/122117-4) and CNPq (304077/77-1 and 306122/88-0),
PRONEX/CNPq.

\end{acknowledgements}

\clearpage

\begin{figure}
 \plotone{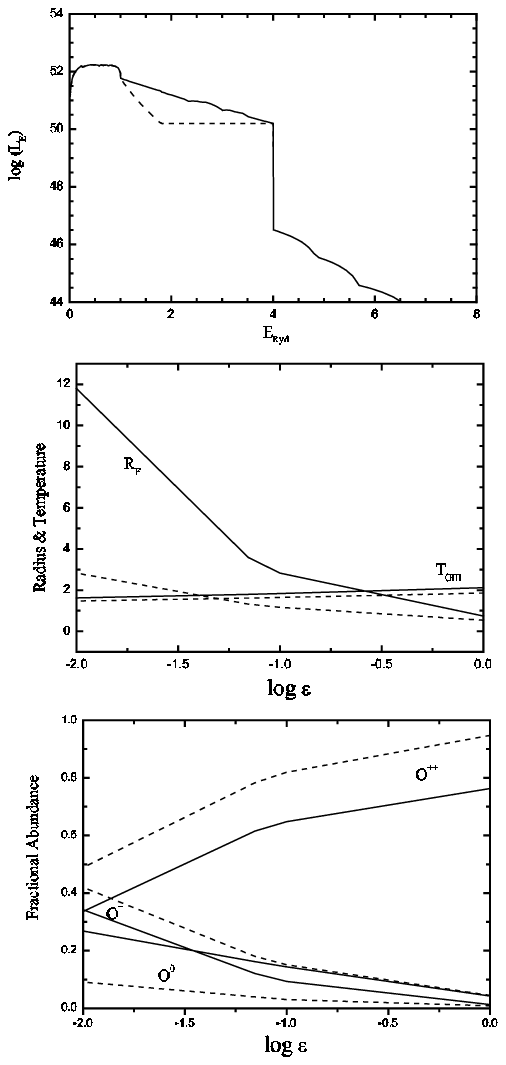}
 \caption{The effect of the filling factor on 1-D models.
 Top-panel: The zero-age stellar cluster ionizing spectrum (solid
 line) and a modified spectrum, with less photons above the
 Lyman limit (dashed line). Middle-panel: the radius of the ionized
 region (R$_F$/10$^{20}$cm$^{-2}$) and the average temperature 
(T$_{OIII}$/10$^4$K) as a function of the filling factor.
 Bottom-panel: the ionization fraction of O$^0$,
  O$^+$ and  O$^{++}$ as a function of the filling factor.
  The solid lines correspond to the results of models with
  the zero-age stellar cluster spectrum and the dashed lines to
  those obtained with the modified spectrum.
}
 \end{figure}

 \begin{figure}
 \plotone{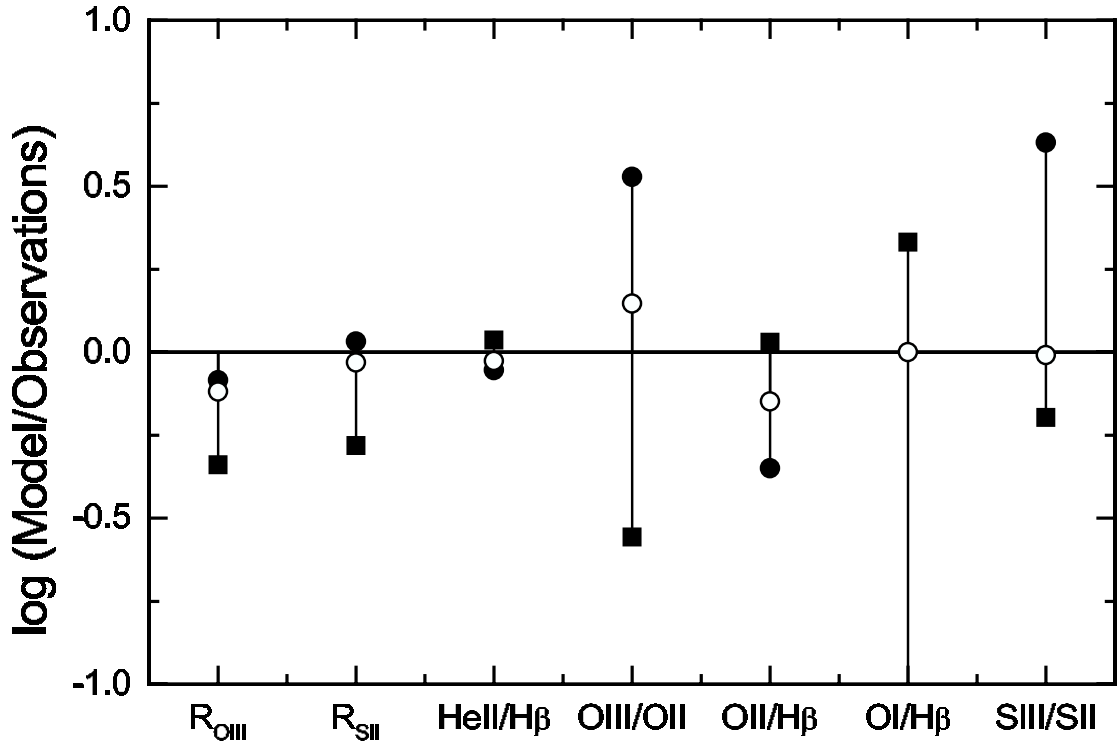}
 \caption{ Comparison of calculated and observed emission-line
        ratios for IZw 18.The results correspond to a homogeneus model
         (filled circle), a condensation
           (filled square) and a composite model (open circle)
            for which 60\% of the H$\beta$ emission
           comes from the dilute homogeneous gas distribution (30 cm$^{-3}$)and
           40 \% from the condensation  (10$^4$  cm$^{-3}$).\label{f1}
}
 \end{figure}

\clearpage

\begin{figure}
\plottwo{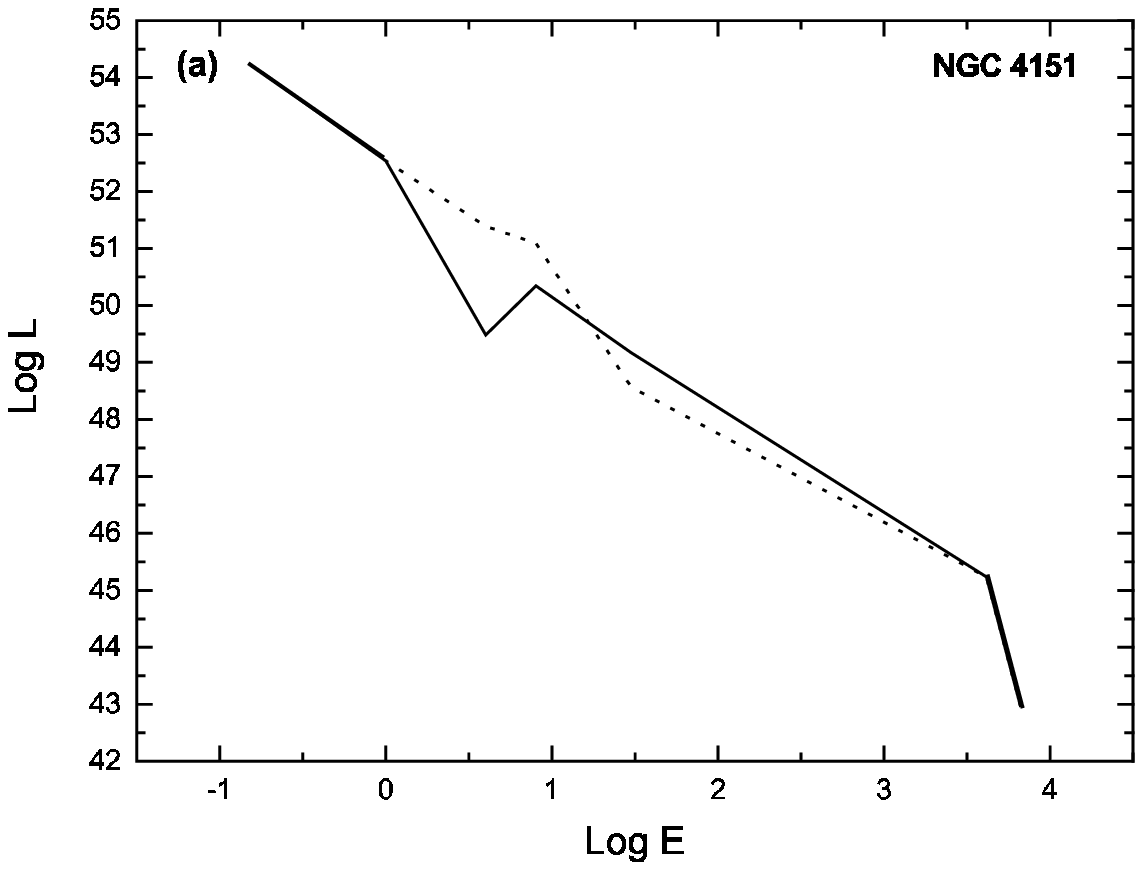}{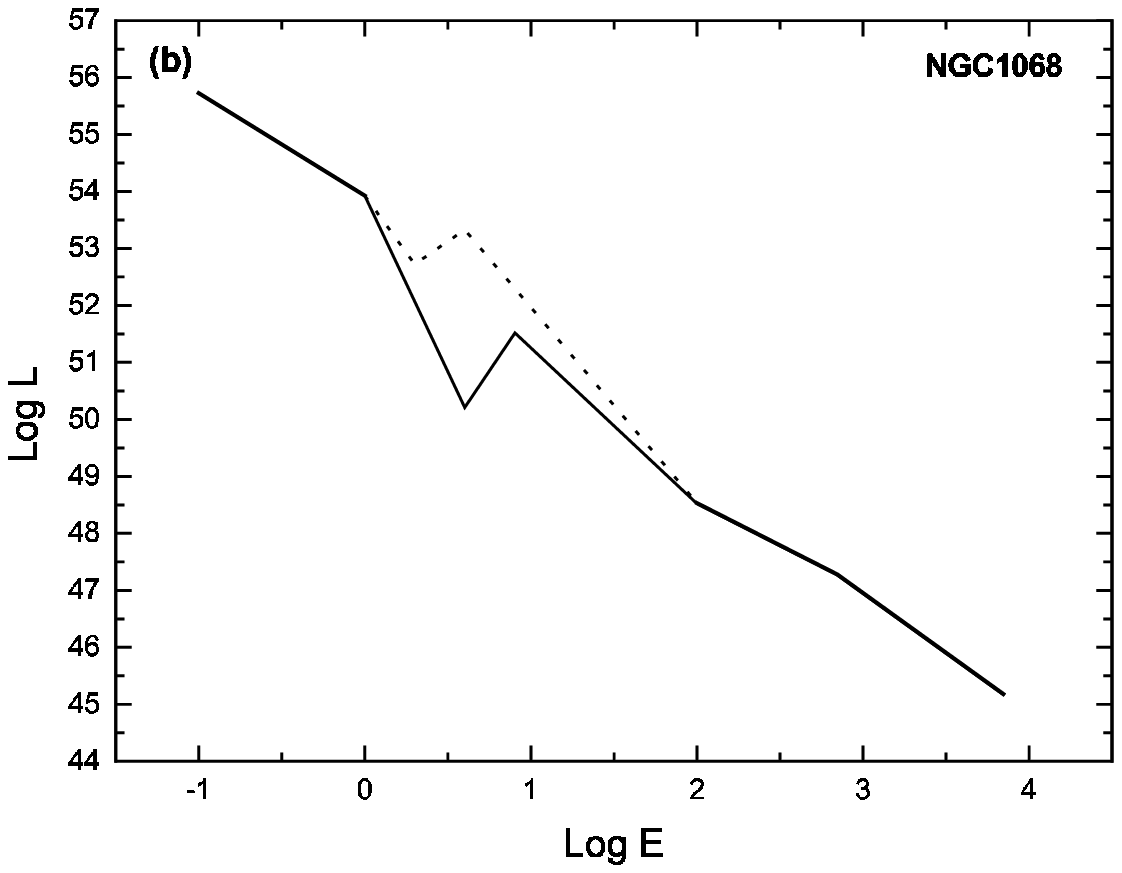}
\caption{(a) Ionizing radiation spectrum  for NGC 4151. The spectral
            luminosity is given in units of photons s$^{-1}$ eV$^{-1}$ and
           the energy in Ryd. The thick solid line corresponds to observations,
            the thin solid line to the spectral shape A used in the best fit model
            of A99, while the dotted line represents a bumpy shape B (see text).
            (b) Ionizing radiation spectrum  for NGC  1068. The notation is
              the same as in Figure 2a.\label{f2}}
\end{figure}

\clearpage

\begin{figure}

\plotone{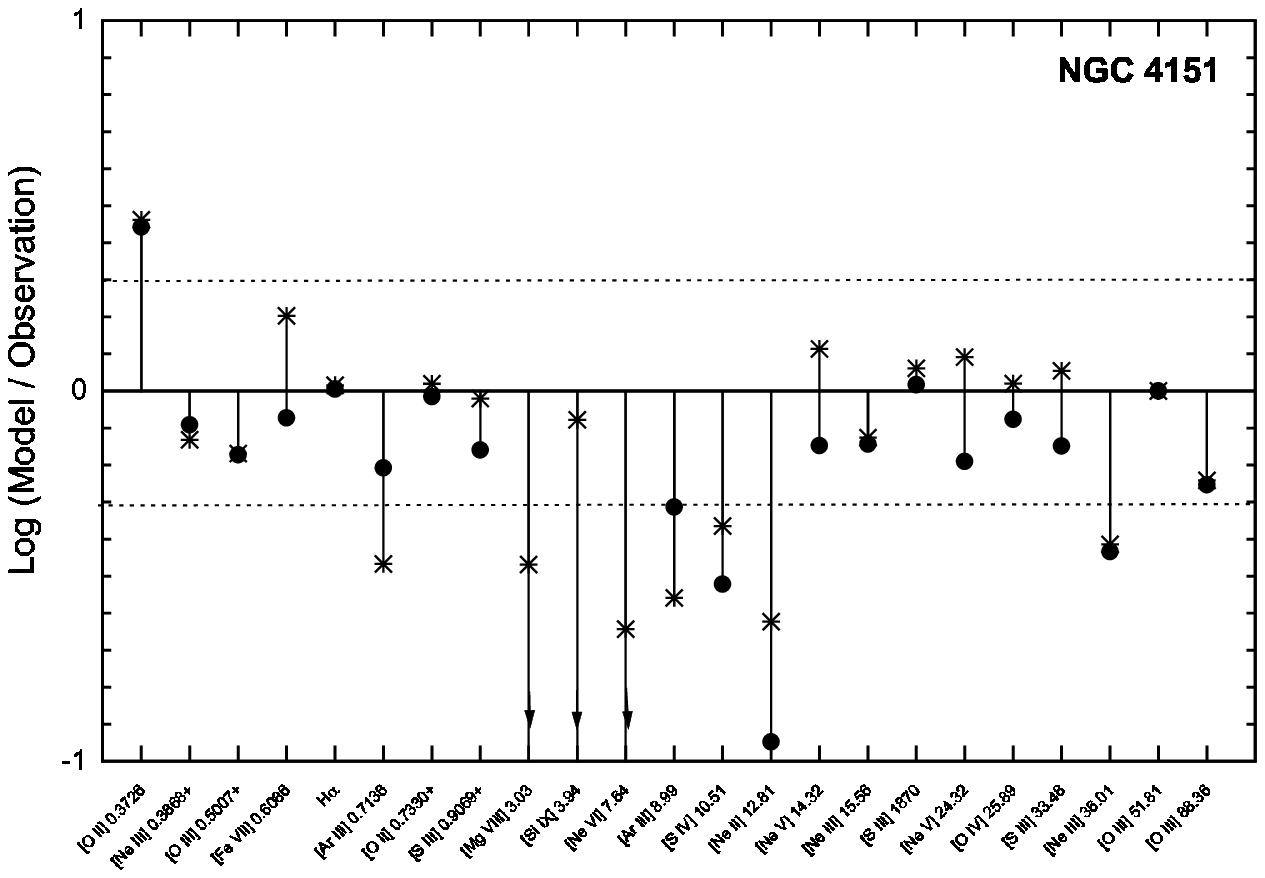}
\caption{Comparison of calculated and observed emission-line ratios for NGC
          4151.
           The results correspond to the single-cloud models using a filling factor
           $\epsilon$ = 0.025 with spectrum A (star) and spectrum B (filled circle).
           The SED of the ionizing radiation are shown in Fig 2a. The emission-lines
           are identified only by the ion in the horizontal axis
           following the order of wavelengths as in Table 1a.\label{f3}}
\end{figure}

\clearpage

\begin{figure}
\plotone{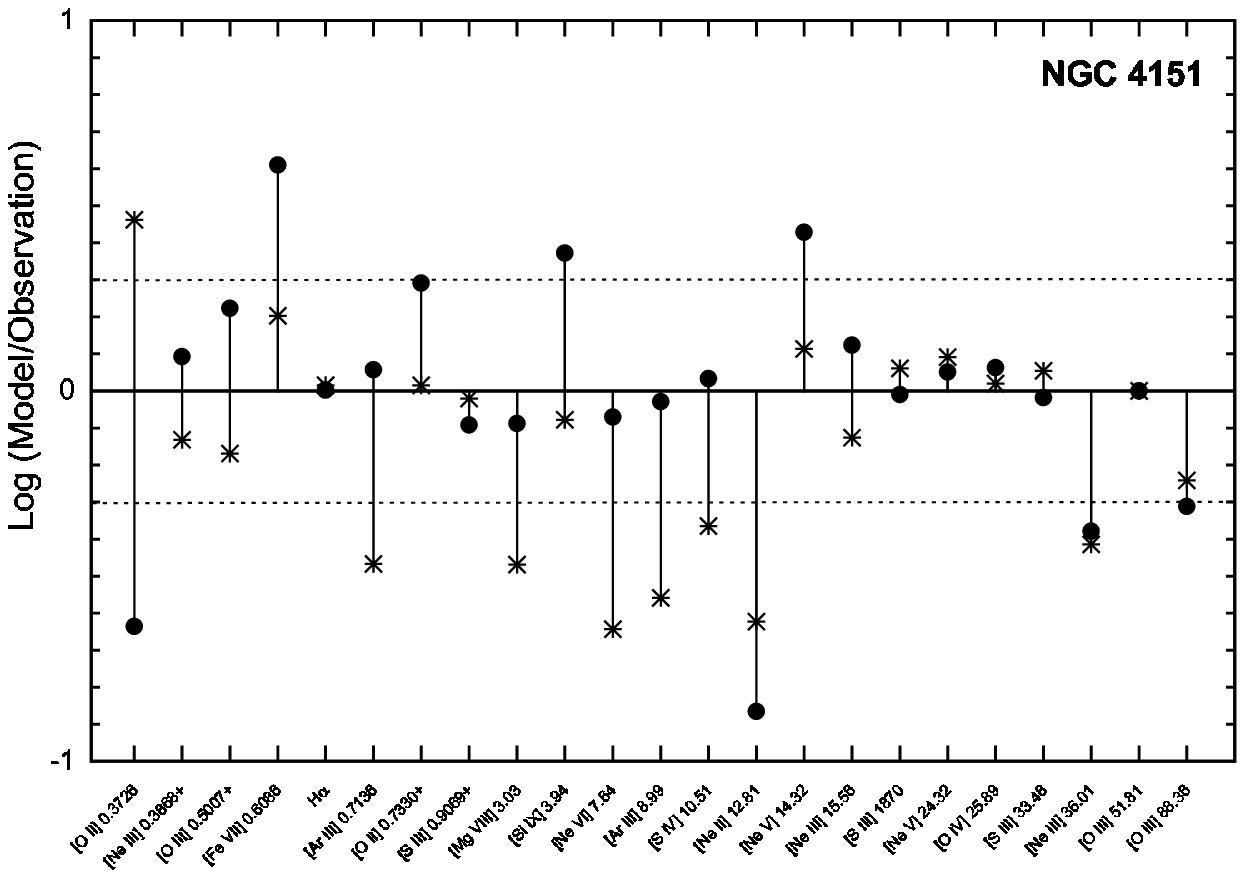}
\caption{ Comparison of calculated and observed emission-line ratios for NGC 4151.
            The results correspond to the single-cloud model using a filling factor
           $\epsilon$ = 0.025 with spectrum A (star) and the multi-cloudmodel
           using spectrum B (filled circle).  The SED of the ionizing radiation
           is shown in Fig 2a. The emission-lines
           are identified only by the ion in the horizontal axis
           following the order of wavelengths as in Table 1a.\label{f4}}
\end{figure}

\clearpage
\begin{figure}
\plotone{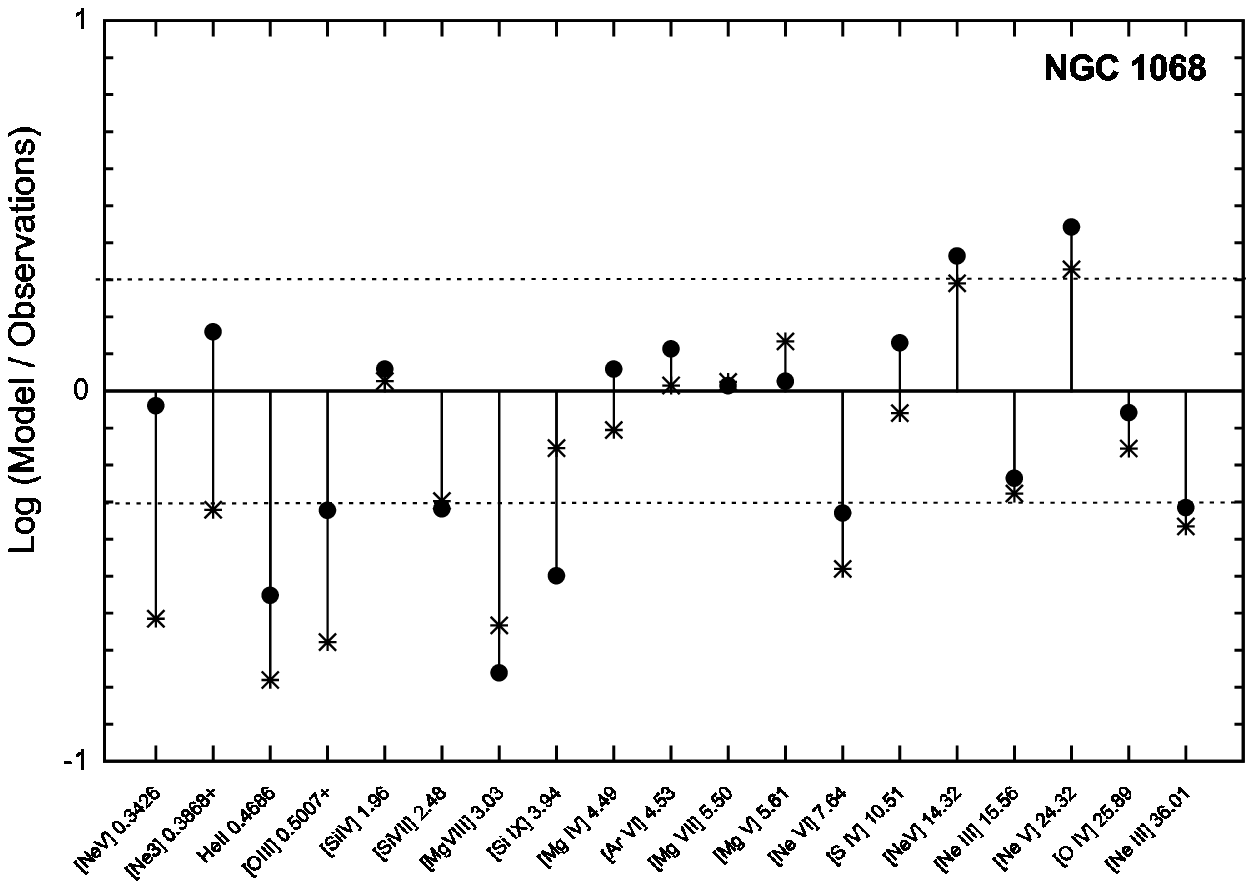}
\caption{Comparison of calculated and observed emission-line ratios for NGC
           1086.
           The results correspond to the single-cloud models using a filling factor
           $\epsilon$ = 0.1 with spectrum A (star) and spectrum B (filedcircle).
           The SED of the ionizing radiation are shown in Fig 2b.The emission-lines
           are identified only by the ion in the horizontal axis
           following the order of wavelengths as in Table 1b.\label{f5}}
\end{figure}

\clearpage

\begin{figure}
\plotone{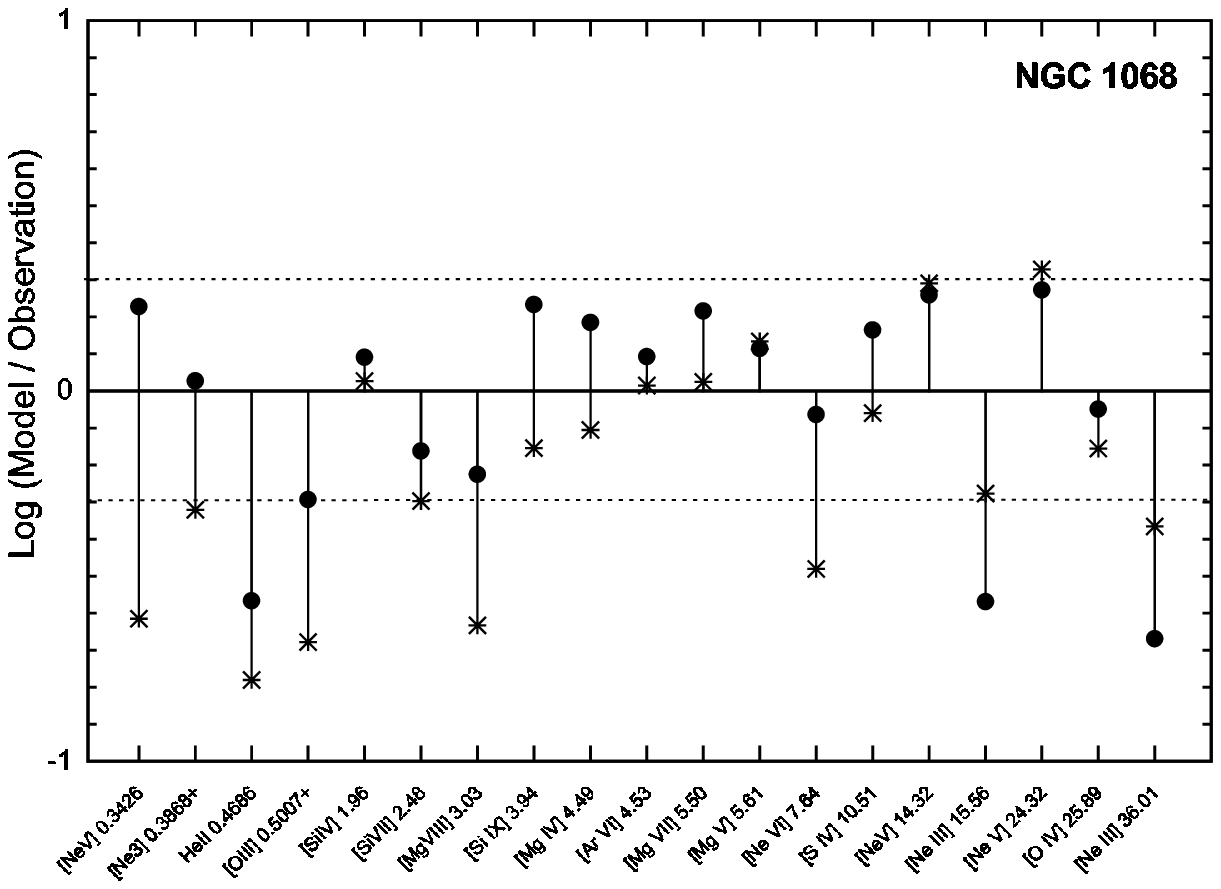}
\caption{Comparison of calculated and observed emission-line ratios for NGC
            1068.
           The results correspond to the single-cloud model using a fillingfactor
           $\epsilon$ = 0.1 with spectrum A (star) and the multi-cloud model
           using spectrum B (filled circle). The SED of the ionizing radiaion
           are shown in Fig 2b.  The emission-lines
           are identified only by the ion in the horizontal axis
           following the order of wavelengths as in Table 1b.\label{f6}}
\end{figure}

\clearpage

\begin{deluxetable}{lr}
\tablecaption{\hskip -7pt a NGC 4151 observed emission-lines\label{1a} }
\tablewidth{0pt}
\tablehead{
\colhead{\bf{Line (\micron)}} & \colhead{ \bf{{I$_{\lambda}$}/{I$_{H\beta}$
}}}\\
}
\startdata
$[$O II]0.3726+ & 2.968\\
$[$Ne III]0.3868+ & 2.196\\
$[$O III]0.5007+ & 13.287\\
$[$Fe VII]0.6086 & 0.113\\
H  I 0.6563 & 2.869\\
$[$Ar III]0.7136 & 0.194\\
$[$O II]0.7330+ & 0.257\\
$[$S  III]0.9069+ & 1.958\\
$[$Mg VIII]3.03 & 0.032\\
$[$Si IX]3.94 & 0.021\\
$[$Ne VI]7.64 & 0.371\\
$[$Ar III]8.99 & 0.103\\
$[$S IV]10.51 & 0.522\\
$[$Ne II]12.81 & 0.541\\
$[$Ne V]14.32 & 0.251\\
$[$Ne III]15.56 & 0.943\\
$[$S III]18.70 & 0.245\\
$[$Ne V]24.32 & 0.253\\
$[$O IV]25.89 & 0.914\\
$[$S III]33.48 & 0.363\\
$[$Ne III]36.01 & 0.157\\
$[$O III]51.81 & 0.460\\
$[$O III]88.36 & 0.302\\

\enddata
\end{deluxetable}

\clearpage

\addtocounter{table}{-1}%
\begin{deluxetable}{lr}
\tablecaption{\hskip -7pt b NGC 1068 observed emission-lines \label{1b}}
\tablewidth{0pt}
\tablehead{
\colhead{\bf{Line (\micron)}} & \colhead{ \bf{{I$_{\lambda}$}/{I$_{H\beta}$
}}}\\
}
\startdata

$[$Ne V]0.3426 & 1.521\\
$[$Ne III]0.3868+ & 1.632\\
He II 0.4686 & 0.381\\
$[$O III]0.5007+ & 15.397\\
$[$Si IV]1.96 & 0.170\\
$[$Si VII]2.48 & 0.175\\
$[$Mg VIII]3.03 & 0.221\\
$[$Si IX]3.94 & 0.096\\
$[$Mg IV]4.49 & 0.142\\
$[$Ar VI]4.53 & 0.280\\
$[$Mg VII]5.50 & 0.237\\
$[$Mg V]5.61 & 0.327\\
$[$Ne VI]7.64 & 1.935\\
$[$S IV]10.51 & 0.996\\
$[$Ne V]14.32 & 1.637\\
$[$Ne III]15.56 & 2.690\\
$[$Ne V]24.32 & 1.159\\
$[$O IV]25.89 & 3.140\\
$[$Ne III]36.01 & 0.279\\

\enddata
\end{deluxetable}

\clearpage

\begin{deluxetable}{lcc}
\tablecaption{Input parameters \label{2}}
\tablewidth{0pt}
\tablehead{
\colhead{}  & \colhead{NGC 4151} & \colhead{ NGC 1068}\\
}
\startdata

U............ & 0.025 & 0.1\\
n............& 10$^{3}$ cm$^{-3}$ & 2 x 10$^{3}$ cm$^{-3}$\\
$\epsilon$ ..........& 6.5 x 10$^{-4}$ & 1 x 10$^{-3}$\\
R$_i$ ..........& 1.46 x 10$^{20}$ cm & 1.82 x 10$^{20}$ cm\\
\enddata
\end{deluxetable}

\clearpage

\begin{deluxetable}{cccc}
\tablecaption{NGC 4151 multi-cloud model \label{3}}
\tablewidth{0pt}
\tablehead{
\colhead{\bf{U}}& \colhead{\bf{n}} & \colhead{ \bf{F$_{H\beta}$}} &
\colhead{ \bf{g$_i$}}\\
\colhead{}     & \colhead{ cm$^{-3}$} & \colhead { 10$^7$ ergs/cm$^{2}$/s}
&\colhead{}  \\
}
\startdata

0.005 & 10$^{3}$ & .002  & 0.082\\
0.005 &10$^{5}$  & 1.90  & 0.869 \\
0.4 &10$^{3}$  &   1.46 & 0.028 \\
0.2 & 10$^{5}$   & 147. & 0.021\\

\enddata
\end{deluxetable}

\clearpage

\begin{deluxetable}{cccc}
\tablecaption{NGC 1068 multi-cloud model\label{4}}
\tablewidth{0pt}
\tablehead{
\colhead{\bf{U}}& \colhead{\bf{n}} & \colhead{\bf{F$_{H\beta}$}}&
\colhead{\bf{g$_i$}}\\
\colhead{}    &\colhead{cm$^{-3}$} & \colhead{ ergs/cm$^{2}$/s} &
\colhead{
}  \\
}
\startdata

0.005 & 10$^{3}$ & .043  & 0.549  \\
0.1 & 10$^{4}$  &  .118  & 0.237\\
0.1 &10$^{3}$   &  1.16 & 0.008\\
0.01 &10$^{5}$  &  10.3  & 0.207  \\

\enddata
\end{deluxetable}

\end{document}